\documentclass[aps,prl,twocolumn,floatfix,reprint]{revtex4}
\usepackage{amsmath,amsbsy,amsfonts,epsfig,bm}

\usepackage{xr}
\newcommand{\D}{\mathrm{d}}

\newcommand{\minttwo}[3]{\int_{#2}^{#3}\! \D #1 \, }

\newcommand{\subfigimg}[3][,]{%
  \setbox1=\hbox{\includegraphics[#1]{#3}}
  \leavevmode\rlap{\usebox1}
  \rlap{\hspace*{0pt}\raisebox{\dimexpr\ht1-9.\baselineskip}{#2}}
  \phantom{\usebox1}
}

\newcommand{\mytitle}{Long-range dispersion forces between molecules subject to ultra-short optical pulses from 
ab initio calculations}

\newcommand{\myauthors}{
\author{Micael~J.~T.~Oliveira}
\altaffiliation[Present address: ]{Max Planck Institute for the Structure and Dynamics of Matter, Luruper Chaussee 149, 22761 Hamburg, Germany}
\affiliation{nanomat/QMAT/CESAM, University of Li\`ege, B-4000, Li\`ege, Belgium}
\affiliation{European Theoretical Spectroscopy Facility http://www.etsf.eu}

\author{Ksenia~Komarova}
\affiliation{Theoretical Physical Chemistry, MolSys, University of Li\`ege, B-4000, Li\`ege, Belgium}

\author{Fran\c{c}oise~Remacle}
\affiliation{Theoretical Physical Chemistry, MolSys, University of Li\`ege, B-4000, Li\`ege, Belgium}

\author{Matthieu~J.~Verstraete}
\email{matthieu.verstraete@ulg.ac.be}
\affiliation{nanomat/QMAT/CESAM, all\'ee du 6 ao\^ut 19 B5a, University of Li\`ege, B-4000, Li\`ege, Belgium}
\affiliation{European Theoretical Spectroscopy Facility http://www.etsf.eu}
}

\begin{document}

\title{\mytitle{}}

\myauthors{}

\begin{abstract}
London-van der Waals dispersion forces are a fundamental component of condensed matter systems, biological processes, and self-assembly. 
In this letter we propose a method to calculate the C$_6$ coefficients that characterize  dispersion  forces in the non-retarded regime 
for molecules in a coherent superposition of excited states. Several ultrafast femtosecond pump-probe schemes are investigated.
We apply the method to LiH molecules and show that their London-van der Waals interaction can change
dramatically after the interaction with the pump pulse. The pulse modulates the C$_6$ coefficients, and the interplay
between polarization, orientation of the molecules, and the dipole fields gives rise to a rich
variety of combinations.
\end{abstract}

\maketitle

The London-van der Waals dispersion forces arising from instantaneously induced dipoles in molecules are a key
ingredient in a wide range of phenomena in physics, chemistry, and biology. Among these, one finds the condensation of 
non-polar gases, the self-assembly of nanostructures, and the folding and dynamics of proteins~\cite{israelachvili11}.
Therefore, the ability to control and manipulate dispersion forces between atoms and molecules is of great importance. 
Because those dispersion interactions depend crucially on the electronic properties of the molecular systems, a simple 
route to achieve this would consist in manipulating their electronic states. Up to now, the vast majority of 
experimental and first-principles studies of dispersion forces focused interacting systems in their electronic ground-state. However, experimental measurements have been made of the van der Waals (vdW)
interaction between Rydberg atoms~\cite{bvclb13:263210} and the problem of calculating the interaction between excited 
atoms or molecules has attracted strong attention recently~\cite{pt95:3660,sherkunov05:052703,dgl15:033201,bprb16:012705}. 

The recent development of ultra-short optical pulses has given researchers unprecedented control over electronic 
degrees of freedom~\cite{vrakking14:2775}. These pulses, tailored in their frequency and envelope, allow the generation 
of a strongly out of equilibrium population of electronic states. Since the generation of the electron wave-packet 
happens before any significant movement of the atomic nuclei, this could be used to steer 
chemical reactions~\cite{rl06:6793}.

In this Letter, we show for the first time how the London interaction coefficients between molecules can be modified 
by an ultrafast laser pump pulse. We describe our method, which employs either standard quantum chemistry or time-dependent 
density functional theory (TD-DFT) to compute the excited states. We show how this method can be used to control the dispersion interaction between LiH 
molecules subject to pulses of Infrared (IR) or Ultraviolet (UV) light. We find that dispersion coefficients
can increase by orders of magnitude in a superposition of states, and even become negative.

There are many methods available to solve the time-dependent Schr\"odinger equation describing the interaction of a 
molecule with a pump pulse. One option is to do so numerically in a basis of multi-determinant field-free 
electronic states that diagonalize the stationary electronic Hamiltonian. The eigenvalues and wavefunctions of the 
field-free electronic states can be obtained using any suitable electronic structure method: configuration interaction 
(CI)~\cite{kks05:074105,ss12:7161,rn12:064104,bsm14:062508}, coupled-cluster~\cite{scs11:4678,ss11:11832}, 
CAS-SCF~\cite{er81:362,joy81:5802,roos87:399,werner87:1,mgvlr11:8331}, etc. As an alternative, one could use methods 
with explicit propagation of  orbitals in time, such as TD-DFT~\cite{mmngr2012}, TD-CI~\cite{rgs06:043420}, or 
TD-CAS~\cite{si13:023402,mm13:062511,rnl07:183902,nrl08:025019}.

The London interaction energy between two molecules $A$ and $B$ in the non-retarded regime can be 
expanded in terms of the inter-molecular distance $R$.
The coefficients of the expansion are known as Hamacker constants~\cite{hamaker37:1058}, and the leading term $-C_6^\mathrm{AB}/R^6$  
can be expressed using the dynamical polarizability tensors $\boldsymbol{\alpha}$ 
of the molecules, at imaginary frequencies $i\omega$~\cite{mclachlan63:387}.
The polarizability $\boldsymbol{\alpha}$ can be obtained either using linear response theory as a sum over states
involving the transition dipole moments and the energy differences. Alternatively, it can be obtained
by solving explicitly the time-dependent equation for the molecule with an impulsive perturbation.
The polarizability is then obtained from the induced dipole moment as a function of time~\cite{slrml12:139}.

The case of the polarizability of a molecule in which one or more excited states are populated is a less 
studied one. This problem is relevant in the context of transient spectroscopy, where the interaction of a molecule with a probe 
is preceded by the application of a pump pulse. As in  the ground-state case, the dynamical polarizabilities 
can be either calculated through explicit real-time propagation of 
the time-dependent equations~\cite{gbcwr13:1363}, or expressed as a sum over states~\cite{ps15:033416}. 

In this work we have chosen to solve the time-dependent Schr\"odinger equation describing the interaction of the system
with the pump pulse numerically in the basis of the
multi-determinant field-free electronic states, and to compute the polarizabilities at any given 
time-delay $T$ after the pump pulse as a sum over states~\cite{ps15:033416}:
\begin{multline}
 \label{eq:sos}
 \alpha(i\omega,T) = \sum_{\alpha\beta\gamma} c^*_\alpha(T) c_\beta(T) d_{\alpha\gamma} d_{\gamma\beta} \\
 \times \left(\frac{1}{i(\omega + \eta) - E_\gamma + E_\alpha} - \frac{1}{i(\omega +\eta) + E_\gamma-E_\beta} 
\right)\,,
\end{multline}
where $c$ are the expansion coefficients of the coherent superposition of states, $d$ are the transition dipole 
moments, and $E$ are the eigenenergies of the states. The quantity $\eta$ is introduced to avoid the divergence of some 
terms at zero frequency, and it corresponds to assuming that the excited states have a finite 
lifetime~\cite{sakurai93}. Calculating $\alpha$ through the Laplace transform of an explicit real-time propagation~\cite{gbcwr13:1363} gives the same results, 
to within numerical accuracy, as using Eq.~\ref{eq:sos}. The choice between the two methods is thus one of convenience.

\begin{figure}[!t]
\centering
\includegraphics[width=0.8\columnwidth]{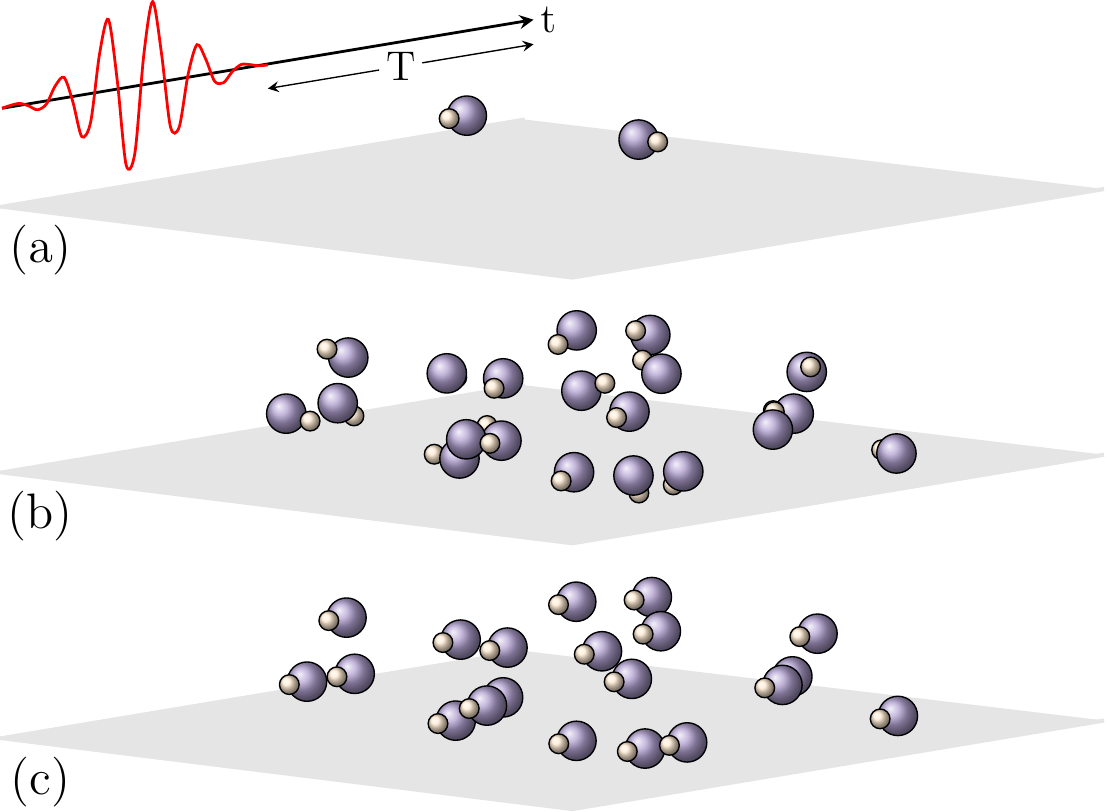}
\caption{Schematic representation of different geometrical arrangements of molecules subject to ultra-short optical 
pump pulses: (a) Two molecules with fixed relative orientations with respect to each other and to the pump pulse. (b) 
Set of randomly oriented molecules. (c) Set of molecules aligned with respect to the pump pulse.}
\label{fig1:exp_setups}
\end{figure}

Before proceeding with the calculation of the Hamacker constants, we consider the possible 
experimental set ups where these can be measured. 
In particular, one needs to consider how the molecules are oriented with respect to each other and to the pump 
pulse, that is, with respect to the lab frame. A schematic representation of the different cases considered is shown in 
Fig.~\ref{fig1:exp_setups}. In the first case, a single measurement is done and the relative orientation of the 
molecules with respect to each other and to the lab frame are known and remain fixed (Fig.~\ref{fig1:exp_setups}(a)). 
This is probably the most difficult case to set up experimentally. The second case considered is a sample of randomly oriented molecules (Fig.~\ref{fig1:exp_setups}(b)). In this case it 
is necessary to perform a rotational average of the Hamacker constants over all possible 
orientations of the molecules with respect to each other and to the lab frame. This yields the 
well known expression:
\begin{equation}
\label{eq:c6}
C_6^\mathrm{AB} = \frac{3}{\pi} \minttwo{\omega}{0}{\infty}\left\langle \boldsymbol{\alpha}^\mathrm{A}(i\omega) 
\right\rangle \left\langle \boldsymbol{\alpha}^\mathrm{B}(i\omega)\right\rangle \,,
\end{equation}
where the brackets denote rotational averaging. In the case of molecules in their ground-state one obtains the simple 
analytical expression $\left\langle \boldsymbol{\alpha}(i\omega)\right\rangle = 
\mathrm{Tr}[\boldsymbol{\alpha}(i\omega)]/3$~\cite{andrews04:877}. This expression is not valid in the presence of 
the pump pulse, as the polarizabilities of the molecules will now also depend on their relative 
orientation with respect to the lab frame: molecules with different orientations will interact differently with the pump and 
will thus be driven to a different superposition of states. This means that the rotational average needs to be 
performed numerically to sample all possible relative orientations.

We also consider a gas of molecules which have been aligned (e.g. as in Refs.~\cite{ss03:543,as12:885}) 
prior to the pump pulse (Fig.~\ref{fig1:exp_setups}(c)). 
Their interaction will now only depend on relative positions, and experiments 
yield the corresponding partial average. This leads to a simple analytical expression and only needs to consider 
a single orientation of the molecules in the lab frame.
(Formulae and derivations in the Supplemental Information (SI)).
In all cases we assume that the distance between the molecules is large enough that there is no 
overlap between their electronic densities, but still small enough so that the long-range interaction is not in the 
retarded regime.

As a test system we have chosen the LiH molecule. Heteronuclear diatomics have well-separated excited states 
 which obey simple selection rules.
The $\Sigma$ excited states exhibit different polarities and the $\Pi$ states have a polarized electron density 
far from the internuclear axis~\cite{rnl07:183902,nrl08:025019}. 
Using the pump frequency and polarization we can target specific excited states or superpositions.
The LiH field-free states are determined using CAS-SCF~\cite{er81:362,joy81:5802,roos87:399,werner87:1}, which for LiH is close 
to the full CI limit (details are the same as in Ref.~\cite{omkprv15:2221}). We choose three pump 
pulses, with the same envelope, and different intensities and carrier frequencies, so as to excite coherent
superpositions of states with different features in the beating electronic density, either along the molecular axis 
or with a perpendicular component.
The first pulse has an IR frequency which will populate the 1$\Sigma$ state by a two photon process (3.1 eV) when
the polarization is parallel to the molecular axis.
With a perpendicular polarization the 1$\Pi$ transition is also accessible with two photons, 
given the pulse bandwidth, but it is further from resonance.
The two other pulses carry UV frequencies resonant with the 1$\Sigma$ and 1$\Pi$ states (labeled 1$\Sigma$ and 1$\Pi$, resp.). As for the IR pulse, the character of the populated states 
depends on the pulse polarization having a component parallel ($\Sigma$) or perpendicular ($\Pi$) to the molecular axis.
Note that $\Sigma$ states can also be populated with a perpendicularly polarized pulse, through 
further transitions from populated $\Pi$ states. 
All the pulse intensities are chosen such that no significant populations of excited states occur above the 1$\Pi$ excited state, and we avoid ionizing regimes which would complicate the dynamics.
Detailed pulse parameters are in the SI.

All our calculations were carried out for the equilibrium geometry of the LiH ground-state, as no significant motion 
of the nuclei should occur within the time-delays considered.
When calculating the polarizabilities, we have endowed each electronic wave-packet with the same lifetime $\tau=$ 50 fs. 
This lifetime is large enough to characterize the time evolution of the London interaction, while
processes causing dephasing of the electronic coherence, like luminescence, fragmentation, or non-adiabatic coupling induced by nuclear motion,
should occur on longer time scales~\cite{Ajay2016,Nikodem2016, Nikodem2017}.
We note that the choice of the lifetime can have a significant effect when calculating the polarizabilities:
Terms in Eq.~\ref{eq:sos} where $\gamma=\alpha$ or 
$\gamma=\beta$ will have a divergence when $\omega \rightarrow 0$ and $\eta = 2/\tau \rightarrow 0$ and will 
dominate the integral of Eq.~\ref{eq:c6} when the lifetime is sufficiently large.
Calculations showing the dependence of the C$_6$ on the lifetime are included in the SI.

\begin{figure}[!h]
\centering
\includegraphics[width=1.0\columnwidth]{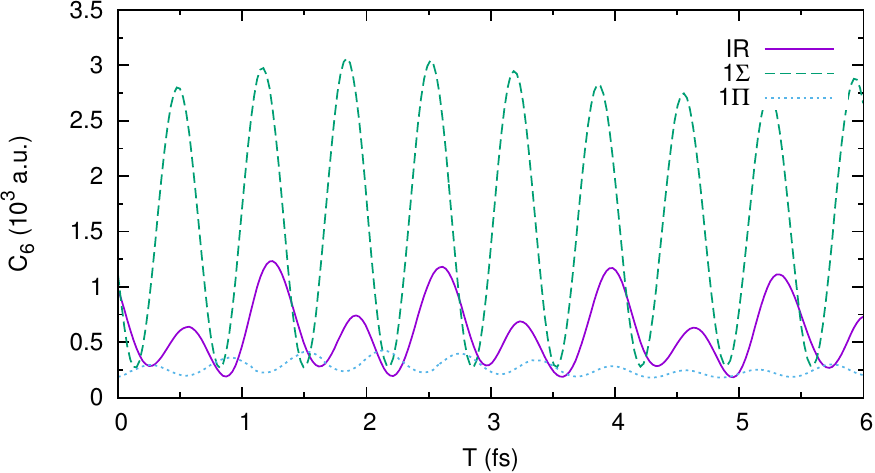}
\caption{Averaged C$_6$ a system of randomly oriented LiH molecules after interacting with an IR optical pump 
pulse and two UV pump pulses resonant with the 1$\Sigma$ and 1$\Pi$ excited states, as a function of the time delay.}
\label{fig2:c6_random}
\end{figure}

The orientation averaged C$_6$, as a function of the time delay, are presented in
Fig.~\ref{fig2:c6_random} for a system of randomly oriented LiH molecules subject 
to the IR pump pulse and to the UV pump pulses resonant with the 1$\Sigma$ and 1$\Pi$ states.
The pump pulse builds a coherent superposition of the electronic states~\cite{rnl07:183902}.
The oscillations as a function of the delay $T$ originate in the expansion coefficients $c_\alpha(T)$: 
after the pump pulse, the populations of the different excited states remain constant, but the coefficients acquire a 
phase $c_\alpha(T)=c_\alpha(0) e^{-iE_\alpha T}$. Because the C$_6$ coefficients depend on the product of polarizabilities,
their dependence on $T$ can be written as a linear combination of terms of the following
form: $e^{-i(E_\alpha - E_\beta + E_\gamma - E_\delta)T}$. In Fig.~\ref{fig2:c6_random} 
the main frequencies in the C$_6$ oscillations are E$_{1\Sigma}-$E$_\mathrm{GS}$ and 2(E$_{1\Sigma}-$E$_\mathrm{GS}$)
for the IR pump pulse (the $\Pi$ state is further from resonance); 2(E$_{1\Sigma}-$E$_\mathrm{GS}$) for the 1$\Sigma$ pulse; 
and E$_{1\Sigma}+$E$_{1\Pi}-2$E$_\mathrm{GS}$ for the 1$\Pi$ pulse.

\begin{figure}[htb]
\centering
\includegraphics[width=\columnwidth]{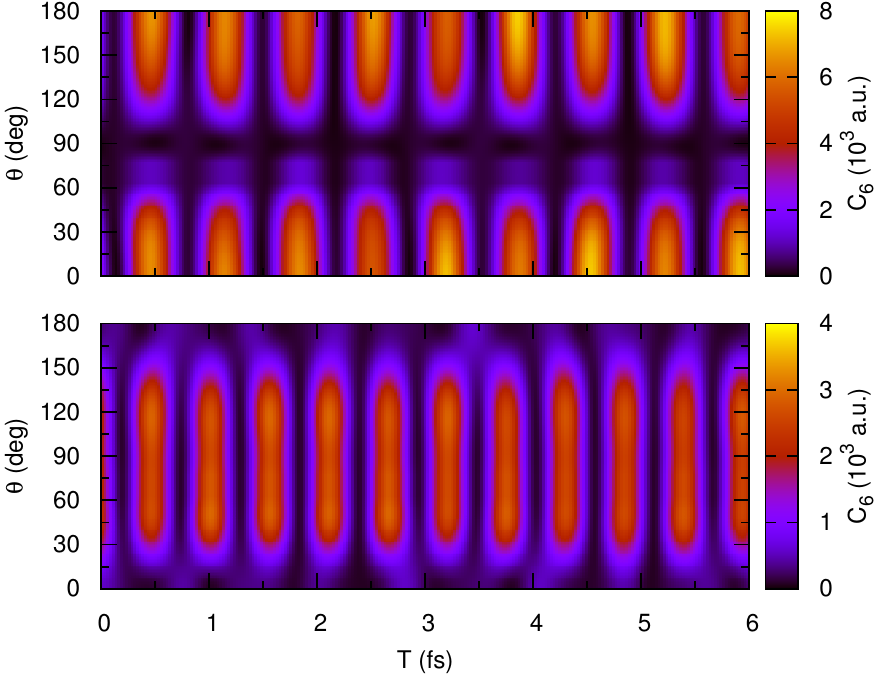}
\caption{Average C$_6$ coefficient of a system of aligned LiH molecules, as a function of delay time and angle 
between pump polarization direction and molecular axis. Upper panel: pump resonant with 1$\Sigma$; Lower panel: resonant with 1$\Pi$.}
\label{fig3:c6_aligned}
\end{figure}

The amplitude of the predicted C$_6$ coefficients can be several orders of magnitude larger
than the ground state value of 243 a.u., and the time averaged value of the C$_6$ is quite different depending on the pump pulse.
We therefore expect a large measurable effect in the vdW forces when the molecules are subject to specific pump pulses.
This configuration (random orientations) is the most representative for gas phase experiments. Even in the pessimistic case of
a short lifetime (tens or hundreds of fs), in light atom molecular dimers the resulting forces should have time to influence
intermolecular distances, vibrational frequencies and other observables. The net classical momentum $F \cdot \delta t$ 
gained is equivalent to that of the ground state London force over many picoseconds. 
Longer lived electronic superpositions will have the double
effect of increasing further the C$_6$ (integrating the divergence further towards $i \omega = 0$) and allowing 
for more time to move atoms subjected to enhanced vdW forces. Experiments by Despre et al.\cite{Despre2015} recently demonstrated sustained coherence in the presence of atomic motion, over at least tens of femtoseconds. 
Recent calculations suggest electronic coherence can survive well over 100 fs in LiH in the presence of nuclear motion~\cite{Nikodem2016}.

\begin{figure*}[tbh]
\centering
\subfigimg[width=0.329\textwidth]{(a)}{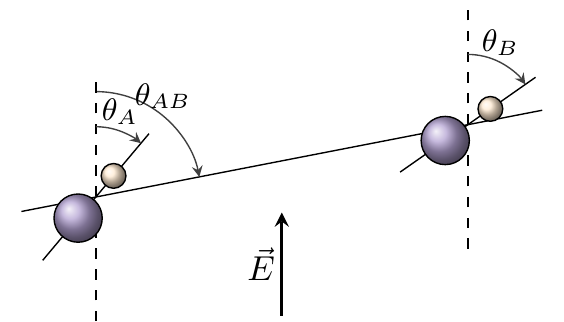}
\subfigimg[width=0.329\textwidth]{(b)}{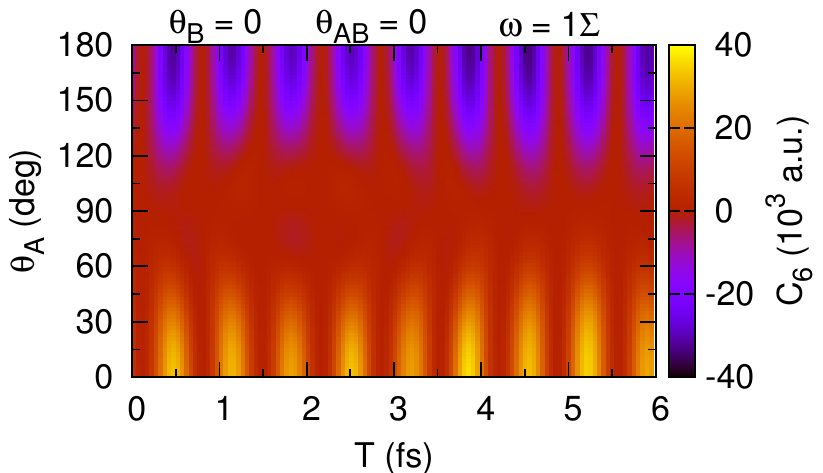}
\subfigimg[width=0.329\textwidth]{(c)}{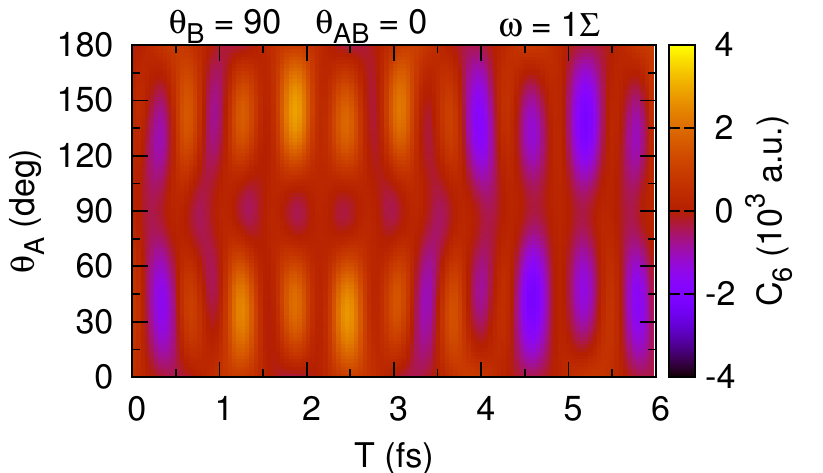}
\subfigimg[width=0.329\textwidth]{(d)}{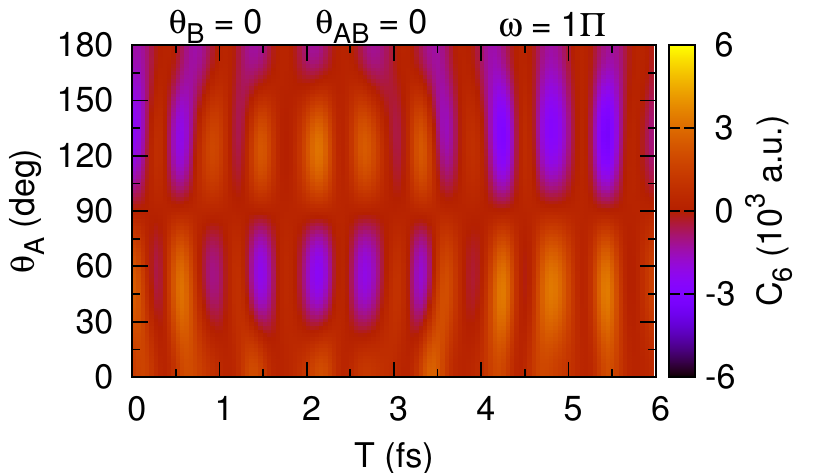}
\subfigimg[width=0.329\textwidth]{(e)}{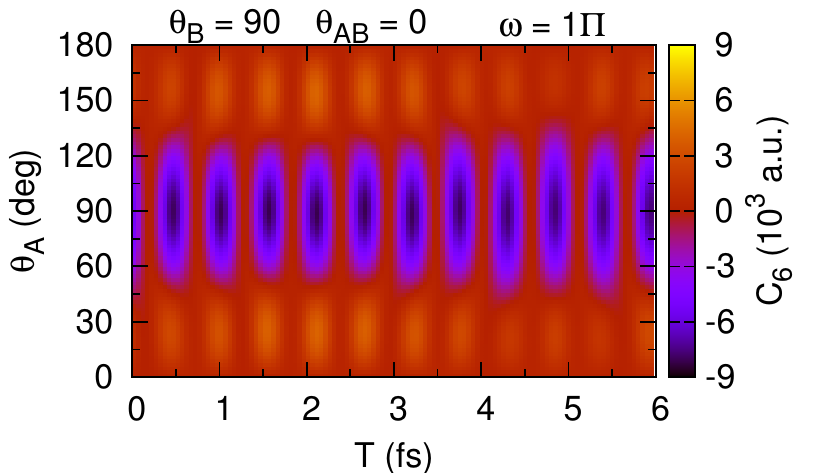}
\subfigimg[width=0.329\textwidth]{(f)}{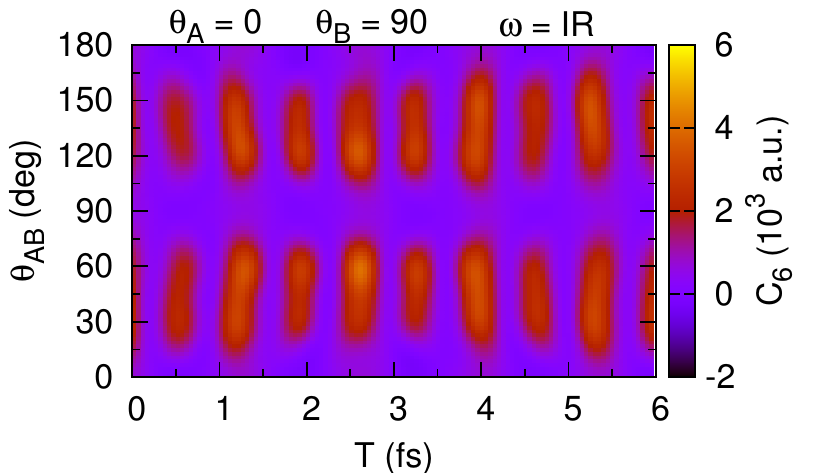}
\caption{C$_6$ values between two LiH molecules as a function of 
 delay time, for selected relative orientations co-planar with the field polarization $\vec{E}$. 
(a) The orientations of the molecules with respect to $\vec{E}$ and to each other are given by the 
angles $\theta_A$, $\theta_B$, and $\theta_{AB}$. 
(b)-(c) Pump resonant with the 1$\Sigma$ state. 
(d)-(e) Pump resonant with the 1$\Pi$ state.
(f) IR  pump which populates both states. 
In (b)-(e) only one molecule is rotated with respect to the lab frame, with the other angles fixed.
In (f) the orientations of both molecules are fixed, while their relative position varies.}
\label{fig4:c6_fixed}
\end{figure*}

As a second example, we consider in Fig.~\ref{fig3:c6_aligned} the C$_6$ of a gas of aligned LiH 
molecules, as a function of the time delay and of the angle between the polarization and the collective molecular 
axis. The amplitude of the oscillations is naturally largest when the population of the targeted 
excited state is maximized. For the 1$\Sigma$ pump pulse, this corresponds to 
$\theta = 0$ and $\theta = 180$, while for the 1$\Pi$ case the pulse polarization 
should be perpendicular to the molecules ($\theta = 90$). This setup is also experimentally within reach,
using a pre-pulse which sets up the molecules in a rotational wave packet~\cite{ss03:543,as12:885}.

In Fig.~\ref{fig4:c6_fixed} we plot C$_6$ values, as a function of the time-delay, between two LiH molecules for 
different relative orientations and pump pulses. The aim is to decompose two main effects that contribute to the 
orientationally averaged results: the nature of the states present in the coherent superposition of electronic states
and the relative orientations of the molecules over which the averages are done.
To simplify the analysis, we only consider cases where the pulse polarization and the two 
molecules are co-planar. The relative orientations can then be specified by three angles in the lab frame set by the 
field polarization: the polar angles of the main molecular axis of each molecule, and the polar angle of the vector from 
one center of mass to the other (Fig.~\ref{fig4:c6_fixed}(a)). We will now decompose the dependencies of the net C$_6$ response
on symmetry, geometry, and frequency resonance.
In Fig.~\ref{fig4:c6_fixed}(b) the C$_6$ values are plotted as a function of the angle 
between the first molecule and the pump pulse.
The population of the 1$\Sigma$ excited state of the second molecule is maximized, with its axis 
parallel to the pump polarization, and a 1$\Sigma$ resonant carrier frequency. As above, the C$_6$ values oscillate as a function of time, but the sign and amplitude 
of the oscillations depend strongly on the orientation of the first molecule. When the molecules are parallel 
($\theta_{A}=0^{\circ}$), they are driven to the same superposition of excited states, with a significant population 
($\sim20\%$) in the 1$\Sigma$ state: the interaction between the molecules is then maximal. When the first molecule is perpendicular
($\theta_A=90^{\circ}$), no $\Sigma$ states are populated and the interaction is much smaller. When the
molecules are anti-parallel ($\theta_A=180^{\circ}$), the interaction is again large, but 
the populations and phases of the wave-packets are different.
Therefore, contrary to the first impression conveyed by Fig.~\ref{fig4:c6_fixed}(b), an average over $\theta_A$, for each time delay, will give a finite value. This also explains 
why the amplitude of the oscillations is smaller (but not zero) in the randomized case of Fig.~\ref{fig2:c6_random} than in 
Fig.~\ref{fig4:c6_fixed}. 
In Fig.~\ref{fig4:c6_fixed}(c) we also keep molecule B fixed, but perpendicular to 
the polarization direction. As the frequency is further from resonance, the 1$\Pi$ state is weakly populated.
The dependence of the C$_6$ values with respect to the orientation of the first molecule is quite different, and 
there are long time beatings. The shorter beatings correspond to twice $E_{1\Sigma}-E_{GS}$
and the longer ones to twice $E_{1\Pi}-E_{1\Sigma}$.
The next two cases (Figs.~\ref{fig4:c6_fixed}(d)-(e)) are similar, but populating the 1$\Pi$ instead of the 1$\Sigma$ state. 
This 1$\Pi$ population is maximized when the molecule is perpendicular to the pump polarization.
In the last panel (Fig.~\ref{fig4:c6_fixed}(f)) we show a case where the IR pulse 
populates the 1$\Sigma$ state of molecule A 
($\theta_A=0^{\circ}$) and the 1$\Pi$ state of molecule B ($\theta_B=90^{\circ}$).
The C$_6$ shows a maximum for relative orientation of the molecules at angles of 45 and 135$^{\circ}$.

In this Letter we present a method to calculate the C$_6$ coefficients of molecules subject to 
ultra-short laser pulses. Using LiH molecules as a test case, we show that, by driving them to an 
appropriate superposition of excited states, it is possible to considerably increase their London-van der Waals interaction.
The origin of the increase is clear in the integral form of the C$_6$ for partially occupied states, and should carry over to other
molecules. A denser set of excited states in other molecules may be detrimental if it induces quick decoherence,
or beneficial if more transitions participate in the buildup of the interaction.
The effects shown here should be measurable experimentally. 
We have addressed several experimental configurations (geometries, molecule alignments, laser frequencies), 
and our results are quite robust. As in any pump-probe experiment, electronic coherence is key, and we 
use conservative hypotheses for the wave packet lifetime. The efficiency of experimental probes
of the modified C$_6$ will depend on their sensitivity to atomic positions or (better) acquired momentum.
We have finally shown that it is possible to change selectively the interaction of molecules 
that have specific relative orientations. This could be exploited to control the self assembly of 
nanostructures, or to steer chemical reactions including steric or docking effects.

This work was funded by the Belgian Fonds de la Recherche Fondamentale Collective, through project number 2.4545.12 
``Control of attosecond dynamics: applications to molecular reactivity''. F.~R. acknowledges support from the Fonds
National de la Recherche Scientifique (Belgium) and partial support by the U.S. Department of Energy (DOE), Office of 
Science, Basic Energy Sciences (BES) under Award \#DE-SC0012628.
Computer time was provided on Zenobe by the Walloon Region under GA 
1117545, and by PRACE on ARCHER (ThermoSpin and ACEID 3IP RI-312763). The authors thank COST Actions MP1306 
EUSpec and CM1204 XLIC for support.

\bibliography{atto_vdw}

\begin{thebibliography}{39}
\expandafter\ifx\csname natexlab\endcsname\relax\def\natexlab#1{#1}\fi
\expandafter\ifx\csname bibnamefont\endcsname\relax
  \def\bibnamefont#1{#1}\fi
\expandafter\ifx\csname bibfnamefont\endcsname\relax
  \def\bibfnamefont#1{#1}\fi
\expandafter\ifx\csname citenamefont\endcsname\relax
  \def\citenamefont#1{#1}\fi
\expandafter\ifx\csname url\endcsname\relax
  \def\url#1{\texttt{#1}}\fi
\expandafter\ifx\csname urlprefix\endcsname\relax\def\urlprefix{URL }\fi
\providecommand{\bibinfo}[2]{#2}
\providecommand{\eprint}[2][]{\url{#2}}

\bibitem[{\citenamefont{Israelachvili}(2011)}]{israelachvili11}
\bibinfo{author}{\bibfnamefont{J.~N.} \bibnamefont{Israelachvili}},
  \emph{\bibinfo{title}{{Intermolecular and Surface Forces}}}
  (\bibinfo{publisher}{Academic Press}, \bibinfo{year}{2011}),
  \bibinfo{edition}{3rd} ed.

\bibitem[{\citenamefont{B\'{e}guin et~al.}(2013)\citenamefont{B\'{e}guin,
  Vernier, Chicireanu, Lahaye, and Browaeys}}]{bvclb13:263210}
\bibinfo{author}{\bibfnamefont{L.}~\bibnamefont{B\'{e}guin}},
  \bibinfo{author}{\bibfnamefont{A.}~\bibnamefont{Vernier}},
  \bibinfo{author}{\bibfnamefont{R.}~\bibnamefont{Chicireanu}},
  \bibinfo{author}{\bibfnamefont{T.}~\bibnamefont{Lahaye}}, \bibnamefont{and}
  \bibinfo{author}{\bibfnamefont{A.}~\bibnamefont{Browaeys}},
  \bibinfo{journal}{Phys. Rev. Lett.} \textbf{\bibinfo{volume}{110}},
  \bibinfo{pages}{263201} (\bibinfo{year}{2013}).

\bibitem[{\citenamefont{Power and Thirunamachandran}(1995)}]{pt95:3660}
\bibinfo{author}{\bibfnamefont{E.~A.} \bibnamefont{Power}} \bibnamefont{and}
  \bibinfo{author}{\bibfnamefont{T.}~\bibnamefont{Thirunamachandran}},
  \bibinfo{journal}{Phys. Rev. A} \textbf{\bibinfo{volume}{51}},
  \bibinfo{pages}{3660} (\bibinfo{year}{1995}).

\bibitem[{\citenamefont{Sherkunov}(2005)}]{sherkunov05:052703}
\bibinfo{author}{\bibfnamefont{Y.}~\bibnamefont{Sherkunov}},
  \bibinfo{journal}{Phys. Rev. A} \textbf{\bibinfo{volume}{72}},
  \bibinfo{pages}{052703} (\bibinfo{year}{2005}).

\bibitem[{\citenamefont{Donaire et~al.}(2015)\citenamefont{Donaire,
  Gu\'{e}rout, and Lambrecht}}]{dgl15:033201}
\bibinfo{author}{\bibfnamefont{M.}~\bibnamefont{Donaire}},
  \bibinfo{author}{\bibfnamefont{R.}~\bibnamefont{Gu\'{e}rout}},
  \bibnamefont{and}
  \bibinfo{author}{\bibfnamefont{A.}~\bibnamefont{Lambrecht}},
  \bibinfo{journal}{Phys. Rev. Lett.} \textbf{\bibinfo{volume}{115}},
  \bibinfo{pages}{033201} (\bibinfo{year}{2015}).

\bibitem[{\citenamefont{Barcellona et~al.}(2016)\citenamefont{Barcellona,
  Passante, Rizzuto, and Buhmann}}]{bprb16:012705}
\bibinfo{author}{\bibfnamefont{P.}~\bibnamefont{Barcellona}},
  \bibinfo{author}{\bibfnamefont{R.}~\bibnamefont{Passante}},
  \bibinfo{author}{\bibfnamefont{L.}~\bibnamefont{Rizzuto}}, \bibnamefont{and}
  \bibinfo{author}{\bibfnamefont{S.~Y.} \bibnamefont{Buhmann}},
  \bibinfo{journal}{Phys. Rev. A} \textbf{\bibinfo{volume}{94}},
  \bibinfo{pages}{012705} (\bibinfo{year}{2016}).

\bibitem[{\citenamefont{Vrakking}(2014)}]{vrakking14:2775}
\bibinfo{author}{\bibfnamefont{M.~J.~J.} \bibnamefont{Vrakking}},
  \bibinfo{journal}{Phys. Chem. Chem. Phys.} \textbf{\bibinfo{volume}{16}},
  \bibinfo{pages}{2775} (\bibinfo{year}{2014}).

\bibitem[{\citenamefont{Remacle and Levine}(2006)}]{rl06:6793}
\bibinfo{author}{\bibfnamefont{F.}~\bibnamefont{Remacle}} \bibnamefont{and}
  \bibinfo{author}{\bibfnamefont{R.~D.} \bibnamefont{Levine}},
  \bibinfo{journal}{Proc. Natl. Acad. Sci.} \textbf{\bibinfo{volume}{103}},
  \bibinfo{pages}{6793} (\bibinfo{year}{2006}).

\bibitem[{\citenamefont{Krause et~al.}(2005)\citenamefont{Krause, Klamroth, and
  Saalfrank}}]{kks05:074105}
\bibinfo{author}{\bibfnamefont{P.}~\bibnamefont{Krause}},
  \bibinfo{author}{\bibfnamefont{T.}~\bibnamefont{Klamroth}}, \bibnamefont{and}
  \bibinfo{author}{\bibfnamefont{P.}~\bibnamefont{Saalfrank}},
  \bibinfo{journal}{J. Chem. Phys.} \textbf{\bibinfo{volume}{123}},
  \bibinfo{pages}{074105} (\bibinfo{year}{2005}).

\bibitem[{\citenamefont{Sonk and Schlegel}(2012)}]{ss12:7161}
\bibinfo{author}{\bibfnamefont{J.~A.} \bibnamefont{Sonk}} \bibnamefont{and}
  \bibinfo{author}{\bibfnamefont{H.~B.} \bibnamefont{Schlegel}},
  \bibinfo{journal}{J. Phys. Chem. A} \textbf{\bibinfo{volume}{116}},
  \bibinfo{pages}{7161} (\bibinfo{year}{2012}).

\bibitem[{\citenamefont{Raghunathan and Nest}(2012)}]{rn12:064104}
\bibinfo{author}{\bibfnamefont{S.}~\bibnamefont{Raghunathan}} \bibnamefont{and}
  \bibinfo{author}{\bibfnamefont{M.}~\bibnamefont{Nest}}, \bibinfo{journal}{J.
  Chem. Phys.} \textbf{\bibinfo{volume}{136}}, \bibinfo{pages}{064104}
  (\bibinfo{year}{2012}).

\bibitem[{\citenamefont{Bauch et~al.}(2014)\citenamefont{Bauch, S{\o}rensen,
  and Madsen}}]{bsm14:062508}
\bibinfo{author}{\bibfnamefont{S.}~\bibnamefont{Bauch}},
  \bibinfo{author}{\bibfnamefont{L.~K.} \bibnamefont{S{\o}rensen}},
  \bibnamefont{and} \bibinfo{author}{\bibfnamefont{L.~B.}
  \bibnamefont{Madsen}}, \bibinfo{journal}{Phys. Rev. A}
  \textbf{\bibinfo{volume}{90}}, \bibinfo{pages}{062508}
  (\bibinfo{year}{2014}).

\bibitem[{\citenamefont{Sonk et~al.}(2011)\citenamefont{Sonk, Caricato, and
  Schlegel}}]{scs11:4678}
\bibinfo{author}{\bibfnamefont{J.~A.} \bibnamefont{Sonk}},
  \bibinfo{author}{\bibfnamefont{M.}~\bibnamefont{Caricato}}, \bibnamefont{and}
  \bibinfo{author}{\bibfnamefont{H.~B.} \bibnamefont{Schlegel}},
  \bibinfo{journal}{J. Phys. Chem. A} \textbf{\bibinfo{volume}{115}},
  \bibinfo{pages}{4678} (\bibinfo{year}{2011}).

\bibitem[{\citenamefont{Sonk and Schlegel}(2011)}]{ss11:11832}
\bibinfo{author}{\bibfnamefont{J.~A.} \bibnamefont{Sonk}} \bibnamefont{and}
  \bibinfo{author}{\bibfnamefont{H.~B.} \bibnamefont{Schlegel}},
  \bibinfo{journal}{J. Phys. Chem. A} \textbf{\bibinfo{volume}{115}},
  \bibinfo{pages}{11832} (\bibinfo{year}{2011}).

\bibitem[{\citenamefont{Eade and Robb}(1981)}]{er81:362}
\bibinfo{author}{\bibfnamefont{R.~H.~A.} \bibnamefont{Eade}} \bibnamefont{and}
  \bibinfo{author}{\bibfnamefont{M.~A.} \bibnamefont{Robb}},
  \bibinfo{journal}{Chem. Phys. Lett.} \textbf{\bibinfo{volume}{83}},
  \bibinfo{pages}{362} (\bibinfo{year}{1981}).

\bibitem[{\citenamefont{Jorgensen et~al.}(1981)\citenamefont{Jorgensen, Olsen,
  and Yeager}}]{joy81:5802}
\bibinfo{author}{\bibfnamefont{P.}~\bibnamefont{Jorgensen}},
  \bibinfo{author}{\bibfnamefont{J.}~\bibnamefont{Olsen}}, \bibnamefont{and}
  \bibinfo{author}{\bibfnamefont{D.~L.} \bibnamefont{Yeager}},
  \bibinfo{journal}{J. Chem. Phys.} \textbf{\bibinfo{volume}{75}},
  \bibinfo{pages}{5802} (\bibinfo{year}{1981}).

\bibitem[{\citenamefont{Roos}(1987)}]{roos87:399}
\bibinfo{author}{\bibfnamefont{B.~O.} \bibnamefont{Roos}},
  \emph{\bibinfo{title}{The Complete Active Space Self-Consistent Field Method
  and Its Applications in Electronic Structure Calculations}}
  (\bibinfo{publisher}{John Wiley {\&} Sons, Inc.}, \bibinfo{address}{Hoboken,
  NJ, USA}, \bibinfo{year}{1987}), vol.~\bibinfo{volume}{69}, pp.
  \bibinfo{pages}{399--445}.

\bibitem[{\citenamefont{Werner}(1987)}]{werner87:1}
\bibinfo{author}{\bibfnamefont{H.-J.} \bibnamefont{Werner}},
  \emph{\bibinfo{title}{Matrix-Formulated Direct Multiconfiguration
  Self-Consistent Field and Multiconfiguration Reference
  Configuration-Interaction Methods}} (\bibinfo{publisher}{John Wiley {\&}
  Sons, Inc.}, \bibinfo{address}{Hoboken, NJ, USA}, \bibinfo{year}{1987}),
  vol.~\bibinfo{volume}{69}, pp. \bibinfo{pages}{1--62}.

\bibitem[{\citenamefont{Mignolet et~al.}(2011)\citenamefont{Mignolet,
  Gijsbertsen, Vrakking, Levine, and Remacle}}]{mgvlr11:8331}
\bibinfo{author}{\bibfnamefont{B.}~\bibnamefont{Mignolet}},
  \bibinfo{author}{\bibfnamefont{A.}~\bibnamefont{Gijsbertsen}},
  \bibinfo{author}{\bibfnamefont{M.~J.~J.} \bibnamefont{Vrakking}},
  \bibinfo{author}{\bibfnamefont{R.~D.} \bibnamefont{Levine}},
  \bibnamefont{and} \bibinfo{author}{\bibfnamefont{F.}~\bibnamefont{Remacle}},
  \bibinfo{journal}{Phys. Chem. Chem. Phys.} \textbf{\bibinfo{volume}{13}},
  \bibinfo{pages}{8331} (\bibinfo{year}{2011}).

\bibitem[{\citenamefont{Marques et~al.}(2012)\citenamefont{Marques, Maitra,
  Nogueira, Gross, and Rubio}}]{mmngr2012}
\bibinfo{editor}{\bibfnamefont{M.~A.} \bibnamefont{Marques}},
  \bibinfo{editor}{\bibfnamefont{N.~T.} \bibnamefont{Maitra}},
  \bibinfo{editor}{\bibfnamefont{F.~M.} \bibnamefont{Nogueira}},
  \bibinfo{editor}{\bibfnamefont{E.}~\bibnamefont{Gross}}, \bibnamefont{and}
  \bibinfo{editor}{\bibfnamefont{A.}~\bibnamefont{Rubio}}, eds.,
  \emph{\bibinfo{title}{{Fundamentals of Time-Dependent Density Functional
  Theory}}}, vol. \bibinfo{volume}{837} of \emph{\bibinfo{series}{Lecture Notes
  in Physics}} (\bibinfo{publisher}{Springer Verlag},
  \bibinfo{address}{Berlin}, \bibinfo{year}{2012}).

\bibitem[{\citenamefont{Rohringer et~al.}(2006)\citenamefont{Rohringer, Gordon,
  and Santra}}]{rgs06:043420}
\bibinfo{author}{\bibfnamefont{N.}~\bibnamefont{Rohringer}},
  \bibinfo{author}{\bibfnamefont{A.}~\bibnamefont{Gordon}}, \bibnamefont{and}
  \bibinfo{author}{\bibfnamefont{R.}~\bibnamefont{Santra}},
  \bibinfo{journal}{Phys. Rev. A} \textbf{\bibinfo{volume}{74}},
  \bibinfo{pages}{043420} (\bibinfo{year}{2006}).

\bibitem[{\citenamefont{Sato and Ishikawa}(2013)}]{si13:023402}
\bibinfo{author}{\bibfnamefont{T.}~\bibnamefont{Sato}} \bibnamefont{and}
  \bibinfo{author}{\bibfnamefont{K.~L.} \bibnamefont{Ishikawa}},
  \bibinfo{journal}{Phys. Rev. A} \textbf{\bibinfo{volume}{88}},
  \bibinfo{pages}{023402} (\bibinfo{year}{2013}).

\bibitem[{\citenamefont{Miyagi and Madsen}(2014)}]{mm13:062511}
\bibinfo{author}{\bibfnamefont{H.}~\bibnamefont{Miyagi}} \bibnamefont{and}
  \bibinfo{author}{\bibfnamefont{L.~B.} \bibnamefont{Madsen}},
  \bibinfo{journal}{Phys. Rev. A} \textbf{\bibinfo{volume}{89}},
  \bibinfo{pages}{063416} (\bibinfo{year}{2014}), \eprint{1304.5904}.

\bibitem[{\citenamefont{Remacle et~al.}(2007)\citenamefont{Remacle, Nest, and
  Levine}}]{rnl07:183902}
\bibinfo{author}{\bibfnamefont{F.}~\bibnamefont{Remacle}},
  \bibinfo{author}{\bibfnamefont{M.}~\bibnamefont{Nest}}, \bibnamefont{and}
  \bibinfo{author}{\bibfnamefont{R.~D.} \bibnamefont{Levine}},
  \bibinfo{journal}{Phys. Rev. Lett.} \textbf{\bibinfo{volume}{99}},
  \bibinfo{pages}{183902} (\bibinfo{year}{2007}).

\bibitem[{\citenamefont{Nest et~al.}(2008)\citenamefont{Nest, Remacle, and
  Levine}}]{nrl08:025019}
\bibinfo{author}{\bibfnamefont{M.}~\bibnamefont{Nest}},
  \bibinfo{author}{\bibfnamefont{F.}~\bibnamefont{Remacle}}, \bibnamefont{and}
  \bibinfo{author}{\bibfnamefont{R.~D.} \bibnamefont{Levine}},
  \bibinfo{journal}{New J. Phys.} \textbf{\bibinfo{volume}{10}},
  \bibinfo{pages}{025019} (\bibinfo{year}{2008}).

\bibitem[{\citenamefont{Hamaker}(1937)}]{hamaker37:1058}
\bibinfo{author}{\bibfnamefont{H.}~\bibnamefont{Hamaker}},
  \bibinfo{journal}{Physica} \textbf{\bibinfo{volume}{4}},
  \bibinfo{pages}{1058} (\bibinfo{year}{1937}).

\bibitem[{\citenamefont{McLachlan}(1963)}]{mclachlan63:387}
\bibinfo{author}{\bibfnamefont{A.~D.} \bibnamefont{McLachlan}},
  \bibinfo{journal}{Proceedings of the Royal Society A: Mathematical, Physical
  and Engineering Sciences} \textbf{\bibinfo{volume}{271}},
  \bibinfo{pages}{387} (\bibinfo{year}{1963}).

\bibitem[{\citenamefont{Strubbe et~al.}(2012)\citenamefont{Strubbe, Lehtovaara,
  Rubio, Marques, and Louie}}]{slrml12:139}
\bibinfo{author}{\bibfnamefont{D.~A.} \bibnamefont{Strubbe}},
  \bibinfo{author}{\bibfnamefont{L.}~\bibnamefont{Lehtovaara}},
  \bibinfo{author}{\bibfnamefont{A.}~\bibnamefont{Rubio}},
  \bibinfo{author}{\bibfnamefont{M.~A.~L.} \bibnamefont{Marques}},
  \bibnamefont{and} \bibinfo{author}{\bibfnamefont{S.~G.} \bibnamefont{Louie}},
  in \emph{\bibinfo{booktitle}{Fundamentals of Time-Dependent Density
  Functional Theory}}, edited by \bibinfo{editor}{\bibfnamefont{M.~A.~L.}
  \bibnamefont{Marques}}, \bibinfo{editor}{\bibfnamefont{N.~T.}
  \bibnamefont{Maitra}}, \bibinfo{editor}{\bibfnamefont{F.~M.~S.}
  \bibnamefont{Nogueira}}, \bibinfo{editor}{\bibfnamefont{E.~K.~U.}
  \bibnamefont{Gross}}, \bibnamefont{and}
  \bibinfo{editor}{\bibfnamefont{A.}~\bibnamefont{Rubio}}
  (\bibinfo{publisher}{Springer Verlag}, \bibinfo{address}{Berlin},
  \bibinfo{year}{2012}), vol. \bibinfo{volume}{837} of
  \emph{\bibinfo{series}{Lecture Notes in Physics}},
  chap.~\bibinfo{chapter}{7}, pp. \bibinfo{pages}{139--166}.

\bibitem[{\citenamefont{{De Giovannini} et~al.}(2013)\citenamefont{{De
  Giovannini}, Brunetto, Castro, Walkenhorst, and Rubio}}]{gbcwr13:1363}
\bibinfo{author}{\bibfnamefont{U.}~\bibnamefont{{De Giovannini}}},
  \bibinfo{author}{\bibfnamefont{G.}~\bibnamefont{Brunetto}},
  \bibinfo{author}{\bibfnamefont{A.}~\bibnamefont{Castro}},
  \bibinfo{author}{\bibfnamefont{J.}~\bibnamefont{Walkenhorst}},
  \bibnamefont{and} \bibinfo{author}{\bibfnamefont{A.}~\bibnamefont{Rubio}},
  \bibinfo{journal}{ChemPhysChem} \textbf{\bibinfo{volume}{14}},
  \bibinfo{pages}{1363} (\bibinfo{year}{2013}).

\bibitem[{\citenamefont{Perfetto and Stefanucci}(2015)}]{ps15:033416}
\bibinfo{author}{\bibfnamefont{E.}~\bibnamefont{Perfetto}} \bibnamefont{and}
  \bibinfo{author}{\bibfnamefont{G.}~\bibnamefont{Stefanucci}},
  \bibinfo{journal}{Phys. Rev. A} \textbf{\bibinfo{volume}{91}},
  \bibinfo{pages}{033416} (\bibinfo{year}{2015}).

\bibitem[{\citenamefont{Sakurai}(1993)}]{sakurai93}
\bibinfo{author}{\bibfnamefont{J.~J.} \bibnamefont{Sakurai}},
  \emph{\bibinfo{title}{{Modern Quantum Mechanics}}}
  (\bibinfo{publisher}{Addison Wesley}, \bibinfo{year}{1993}),
  \bibinfo{edition}{revised} ed., ISBN \bibinfo{isbn}{0805382917}.

\bibitem[{\citenamefont{Andrews}(2004)}]{andrews04:877}
\bibinfo{author}{\bibfnamefont{S.~S.} \bibnamefont{Andrews}},
  \bibinfo{journal}{J. Chem. Educ.} \textbf{\bibinfo{volume}{81}},
  \bibinfo{pages}{877} (\bibinfo{year}{2004}).

\bibitem[{\citenamefont{Stapelfeldt and Seideman}(2003)}]{ss03:543}
\bibinfo{author}{\bibfnamefont{H.}~\bibnamefont{Stapelfeldt}} \bibnamefont{and}
  \bibinfo{author}{\bibfnamefont{T.}~\bibnamefont{Seideman}},
  \bibinfo{journal}{Rev. Mod. Phys.} \textbf{\bibinfo{volume}{75}},
  \bibinfo{pages}{543} (\bibinfo{year}{2003}).

\bibitem[{\citenamefont{Artamonov and Seideman}(2012)}]{as12:885}
\bibinfo{author}{\bibfnamefont{M.}~\bibnamefont{Artamonov}} \bibnamefont{and}
  \bibinfo{author}{\bibfnamefont{T.}~\bibnamefont{Seideman}},
  \bibinfo{journal}{Mol. Phys.} \textbf{\bibinfo{volume}{110}},
  \bibinfo{pages}{885} (\bibinfo{year}{2012}).

\bibitem[{\citenamefont{Oliveira et~al.}(2015)\citenamefont{Oliveira, Mignolet,
  Kus, Papadopoulos, Remacle, and Verstraete}}]{omkprv15:2221}
\bibinfo{author}{\bibfnamefont{M.~J.~T.} \bibnamefont{Oliveira}},
  \bibinfo{author}{\bibfnamefont{B.}~\bibnamefont{Mignolet}},
  \bibinfo{author}{\bibfnamefont{T.}~\bibnamefont{Kus}},
  \bibinfo{author}{\bibfnamefont{T.~A.} \bibnamefont{Papadopoulos}},
  \bibinfo{author}{\bibfnamefont{F.}~\bibnamefont{Remacle}}, \bibnamefont{and}
  \bibinfo{author}{\bibfnamefont{M.~J.} \bibnamefont{Verstraete}},
  \bibinfo{journal}{J. Chem. Theory Comput.} \textbf{\bibinfo{volume}{11}},
  \bibinfo{pages}{2221} (\bibinfo{year}{2015}).

\bibitem[{\citenamefont{Ajay et~al.}(2016)\citenamefont{Ajay, {\v{S}}mydke,
  Remacle, and Levine}}]{Ajay2016}
\bibinfo{author}{\bibfnamefont{J.}~\bibnamefont{Ajay}},
  \bibinfo{author}{\bibfnamefont{J.}~\bibnamefont{{\v{S}}mydke}},
  \bibinfo{author}{\bibfnamefont{F.}~\bibnamefont{Remacle}}, \bibnamefont{and}
  \bibinfo{author}{\bibfnamefont{R.~D.} \bibnamefont{Levine}},
  \bibinfo{journal}{J. Phys. Chem. A} \textbf{\bibinfo{volume}{120}},
  \bibinfo{pages}{3335} (\bibinfo{year}{2016}).

\bibitem[{\citenamefont{Nikodem et~al.}(2016)\citenamefont{Nikodem, Levine, and
  Remacle}}]{Nikodem2016}
\bibinfo{author}{\bibfnamefont{A.}~\bibnamefont{Nikodem}},
  \bibinfo{author}{\bibfnamefont{R.~D.} \bibnamefont{Levine}},
  \bibnamefont{and} \bibinfo{author}{\bibfnamefont{F.}~\bibnamefont{Remacle}},
  \bibinfo{journal}{J. Phys. Chem. A} \textbf{\bibinfo{volume}{120}},
  \bibinfo{pages}{3343} (\bibinfo{year}{2016}).

\bibitem[{\citenamefont{Nikodem et~al.}(2017)\citenamefont{Nikodem, Levine, and
  Remacle}}]{Nikodem2017}
\bibinfo{author}{\bibfnamefont{A.}~\bibnamefont{Nikodem}},
  \bibinfo{author}{\bibfnamefont{R.~D.} \bibnamefont{Levine}},
  \bibnamefont{and} \bibinfo{author}{\bibfnamefont{F.}~\bibnamefont{Remacle}},
  \bibinfo{journal}{Physical Review A} \textbf{\bibinfo{volume}{95}},
  \bibinfo{pages}{053404} (\bibinfo{year}{2017}).

\bibitem[{\citenamefont{Despr{\'{e}} et~al.}(2015)\citenamefont{Despr{\'{e}},
  Marciniak, Loriot, Galbraith, Rouz{\'{e}}e, Vrakking, L{\'{e}}pine, and
  Kuleff}}]{Despre2015}
\bibinfo{author}{\bibfnamefont{V.}~\bibnamefont{Despr{\'{e}}}},
  \bibinfo{author}{\bibfnamefont{A.}~\bibnamefont{Marciniak}},
  \bibinfo{author}{\bibfnamefont{V.}~\bibnamefont{Loriot}},
  \bibinfo{author}{\bibfnamefont{M.~C.~E.} \bibnamefont{Galbraith}},
  \bibinfo{author}{\bibfnamefont{A.}~\bibnamefont{Rouz{\'{e}}e}},
  \bibinfo{author}{\bibfnamefont{M.~J.~J.} \bibnamefont{Vrakking}},
  \bibinfo{author}{\bibfnamefont{F.}~\bibnamefont{L{\'{e}}pine}},
  \bibnamefont{and} \bibinfo{author}{\bibfnamefont{A.~I.}
  \bibnamefont{Kuleff}}, \bibinfo{journal}{J. Phys. Chem. Lett.}
  \textbf{\bibinfo{volume}{6}}, \bibinfo{pages}{426} (\bibinfo{year}{2015}).

\end{thebibliography}

\end{document}